\documentclass[aps,pra,twocolumn,showpacs,superscriptaddress,longbibliography,nofootinbib]{revtex4-1}
\usepackage[utf8]{inputenc}
\usepackage[margin=0.8in]{geometry}
\usepackage{graphicx}
\usepackage{float}
\usepackage{amsmath}
\usepackage{braket}
\usepackage{hyperref}

\newcommand{\weblink}[1]{\href{#1}{#1}}
\newcommand{\figurewidth}{0.48\textwidth}
\begin{document}
\title{Structural and vibrational properties of lithium under ambient conditions within density functional theory}
\author{Michael Hutcheon}
 \email{mjh261@cam.ac.uk}
 \affiliation{Theory of Condensed Matter Group,
  Cavendish Laboratory, J.~J.~Thomson Avenue, Cambridge CB3 0HE,
  United Kingdom} 
\author{Richard Needs}
 \email{rn11@cam.ac.uk}
 \affiliation{Theory of Condensed Matter Group,
  Cavendish Laboratory, J.~J.~Thomson Avenue, Cambridge CB3 0HE,
  United Kingdom} 

\date{\today}

\begin{abstract}
We apply a general first-principles approach to derive the phase diagram of metallic Lithium at ambient pressure between 0 and 350 K, including identification of candidate phases. We use \textit{ab initio} random structure searching (AIRSS) to identify competing phases and supplement the results with calculations of vibrational properties and relevant derived neutron diffraction patterns. Strong quantum nuclear effects are present, prompting a careful treatment of vibrations. We directly map the Born-Oppenheimer surface of Li for the first time, allowing the extention of the normal quasi-harmonic treatment of vibrations to a ``quasi-anharmonic'' approach, where the effects of anharmonicity are included. The Gibbs free energies of the FCC, BCC, HCP and 9R phases are derived using a variety of equations of state. We find that the anharmonic contribution to the Gibbs free energies of Li phases is of the same order as the differences between phases. Anharmonicity also makes a noticeable difference to the 0 K phonon dispersion of the BCC phase, with the largest difference at the  N-point phonon. The ordering of phase transitions that we find agrees with the calculations of Ackland \textit{et al.} [Science 356, 1254–1259 (2017)], even when anharmonic effects are included, suggesting that a quasi-harmonic treatment is sufficient for correct phase behaviour. We show explicitly that the Martensitic phase transition from close-packed to BCC lithium upon heating is driven by entropic contributions to the phonon free energy.

\end{abstract}

\maketitle
\section{Introduction}

Lithium is the lightest simple metal in the periodic table. It has two stable isotopes, Li-7 (which we focus on in this work) and the much rarer Li-6, both of which exhibit a Martensitic (diffusionless) phase transition from BCC to a close-packed mixed phase upon cooling \cite{lithium_new_groundstate, li7_martensite, quantum_effects_in_lithium, 9r_cycling}. Such phase transitions are of significant practical importance, for example in the martensite-austenite transition in steel, upon which many thermomechanical treatments and alloying methods are based \cite{steel_martensite}. Martensitic transformations are also crucial to phenomena such as shape-memory of materials \cite{shape-memory_materials}. Li is the simplest metal that undergoes such a transformation and, because it is such a light element, quantum-nuclear effects make a signifigant contribution to the free energies \cite{ashcroft_quantum_solid, bulk_li_thermodynamics, quantum_effects_in_lithium}. In particular, at low temperatures Li exhibits close-packed polytypism, whereby the Martensitic transition results in a mixture of metastable close-packed phases that can be described by hexagonal layers. The formation of such a mixed close-packed Martensitic phase means that Li is a useful model system for understanding transformation kinetics and characterization of mixed phases, especially where vibrational effects are important \cite{lithium_new_groundstate, minimum_energy_transition_paths, local_crystal_structure}. Here we demonstrate a general framework for deriving the phase behaviour of such systems.

One of the most challenging aspects of predicting the structures and stability of Li phases is the small energy differences between candidate phases, which are of the order of a few meV per atom \cite{lithium_new_groundstate, bulk_li_thermodynamics}. This is because the electronic structure is free-electron like and therefore relatively insensitive to the nuclear positions \cite{li_free_electron_like}. The computational results obtained may therefore be extremely sensitive to numerical accuracy and convergence parameters. We have calculated well-converged energies for the most important structures of Li, such as the BCC, FCC, HCP and 9R phases. These energies include vibrational effects, and we report both quasi-harmonic and fully anharmonic results.

DFT calculations were performed using the \textsc{castep} \cite{castep} plane-wave pseudopotential code and ultrasoft pseudopotentials. We use the \textit{ab initio} random structure searching approach (AIRSS \cite{AIRSS, airss_perspective, airss_silane}) to derive candidate phases and to study a wider range of relevant structures and the resulting energy scales. The AIRSS approach consists of the construction of pseudo-random initial structures with a set of optional constraints on the symmetry, inter-atomic distances, coordination numbers, structural units (based on the bonding that is indicated by chemical considerations) and the dimensionality of the system, etc. These structures are then relaxed using DFT geometry optimization in order to sample candidate low-energy structures.

\section{Vibrational energies}
\label{sec:vibrations_theory}
Vibrational contributions are of crucial importance to the thermodynamics of Li phases \cite{bulk_li_thermodynamics, lithium_new_groundstate, latt_dyn_dense_li}. A hierarchy of approximations allow a systematic, fully quantum-mechanical treatment of such systems. The Hamiltonian within the \textit{harmonic approximation} is
\begin{equation}
\label{eq:harmonic_harmiltonian_atomic_coords}
    H^{(2)} = \sum_i - \frac{1}{2m_i} \nabla_i^2 + \frac{1}{2} \sum_{i,j} \delta_i \delta_j \frac{\partial^2 V(x)}{\partial x_i \partial x_j}.
\end{equation}
where $x_i$ are the nuclear coordinates, $\delta_i$ are the displacements of the nuclei from their equilibrium positions and $V(x)$ is the static lattice energy. Using a suitable change of variables \cite{anharmonic_phonons}, Eq.\ \ref{eq:harmonic_harmiltonian_atomic_coords} can be rewritten as
\begin{equation}
\label{eq:harmonic_hamiltonian}
    H^{(2)} = \sum_{\mathbf{q},\sigma} -\frac{1}{2} \frac{\partial^2}{\partial p_{\mathbf{q},\sigma}^2} + \frac{1}{2} \omega_{\mathbf{q},\sigma}^2 p_{\mathbf{q},\sigma}^2.
\end{equation}
where $p_{\mathbf{q},\sigma}$ are the \textit{phonon coordinates}. This Hamiltonian corresponds to a series of non-interacting simple harmonic oscillators with frequencies $\omega_{\mathbf{q},\sigma}$. In order to include anharmonic effects, the harmonic potential in Eq.\ \ref{eq:harmonic_hamiltonian} is replaced by the static lattice energy of a given phonon perturbation, $V(\{p_{\mathbf{q},\sigma}\},\beta)$ \cite{anharmonic_phonons}, leading to 
\begin{equation}
  H = \left(\sum_{\mathbf{q},\sigma} -\frac{1}{2} \frac{\partial^2}{\partial p_{\mathbf{q},\sigma}^2}\right) + V(\{p_{\mathbf{q},\sigma}\},\beta).
\end{equation}
where $\beta$ is the inverse temperature, which can affect the static lattice energy through entropic effects (Mermin entropy) and smearing of the electronic Fermi surface. We found these effects to be negligible in our calculations, helped by the use of a dense electronic k-point sampling which provides an accurate resolution of the Fermi surface \cite{supplement}.

We calculate an anharmonic correction to the phonon dispersion at each phonon q-point independently by diagonalizing $H$ in the basis of single phonon states at that q-point, $\{\ket{\mathbf{q},\sigma}\}$,
\begin{equation}
\label{eq:principle_axis_Hamiltonian}
    H_{\sigma, \nu}(\mathbf{q}) = \bra{\mathbf{q},\sigma} H \ket{\mathbf{q},\nu}
\end{equation}
Clearly there is an infinite number of excited states ($\sigma$'s) at each q-point; in order to evaluate the matrix elements we truncate this set to the first 20 exited states, which is more than sufficient for our purposes. The anharmonic potential $E_{el}(\{p_{\mathbf{q},\sigma}\},\beta)$, is mapped using DFT calculations for a discrete set of 31 amplitudes for each mode. The potential is then interpolated to a large number of points (5000) using a quadratic spline. The integral in Eq.\ \ref{eq:principle_axis_Hamiltonian} is then carried out over these points to obtain the matrix elements of the Hamiltonian, which is then diagonalized. The resulting eigenvalues are the anharmonic excitation energies at a particular q-point, giving an anharmonic correction to the phonon dispersion. These excitation energies can then be used to construct an anharmonic phonon free energy at any given temperature (including entropic and zero-point effects) \cite{phonon_entropy_theory, phonons_in_dft}. We find that this method is around 50 times more expensive than traditional harmonic phonon calculations.

Application of a Legendre transform to the free energies provides the Gibbs free energy at a given temperature and pressure via the following minimization:
\begin{equation}
    G(T, P) = \min_V[F(T,V) + PV]
\end{equation}
The thermodynamically most stable phase has the lowest Gibbs free energy at a given temperature and pressure. We perform this minimization by fitting $F(T,V)$ data to a suitable equation of state (see results section \ref{sec:quasi-harmonic_results}). Within this method the effects of thermal expansion and any anharmonic contributions to the equilibrium volume of the system are included.

\section{Results and discussion}
\subsection{Structure searching}
\label{sec:structure_searching}
The FCC, BCC, HCP and 9R structures have all been proposed in the past to explain the experimental data for Li \cite{lithium_new_groundstate,overhauser_Li, alkali_metals_low_temp_xray, barrett_Li}. 
We find that these structures can be recovered very quickly from first principles calculations using \textit{ab initio} random structure searching (AIRSS \cite{AIRSS, airss_perspective, airss_silane}) using only very simple constraints on the initial structures. In particular we constrain the volume per atom to be within 50\% of the known value (based on the density of solid BCC Li) and require that no two atoms are within 1\r{A} of one another (approximately 1/3 of the nearest neighbour distance in BCC Li). The unit cells are generated to contain between 1 and 6 atoms. Once a cell has been generated according to these rules we perform a DFT geometry optimization. This minimizes the electronic free energy and locates a locally stable structure (neglecting expensive-to-calculate vibrational effects). The results of an AIRSS search with these constraints are shown in Table \ref{tab:AIRSS_ht}. We see that it is possible to obtain a good heuristic understanding of the energy landscape, even when the effects of vibrations are neglected. The electronic energies of these phases are extremely similar to one another, with differences on the order of meV/atom. Several other searches were also carried out with additional constraints on the symmetry of the initial structures \cite{supplement}, but this was not found to be useful in this case.

\subsubsection{Close packed structures}
\label{sec:close_packed_structures}
\begin{figure*}
    \vspace{-0.5cm}
    \centering
    \includegraphics[width=\textwidth]{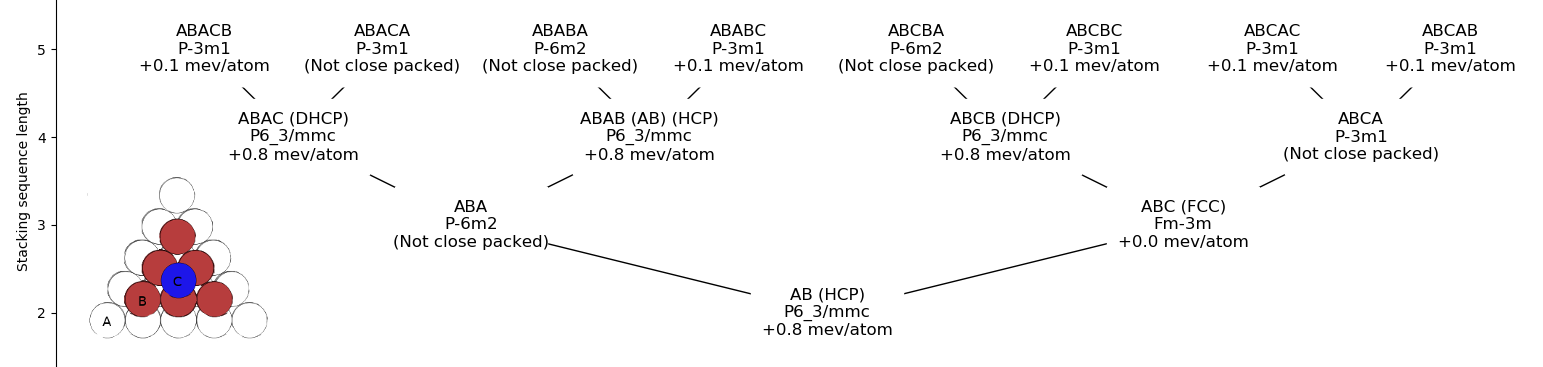}
    \caption{The possible hexagonal-layer stacking sequences from 2 to 5 layers, showing the stacking sequence, the space group and the electronic energy per atom relative to the FCC (ABC stacking) electronic ground state. Note that the energy differences are on the meV scale. The inset is an illustration of the A, B and C type layers. Structures which are not close packed as a result of having the same first and last layer in the sequence are labeled as such.}
    \label{fig:hexagonal_layers}
\end{figure*}

In the previous section we saw that the AIRSS searches quickly recovered the FCC, BCC, HCP and 9R structures. Most of the other structures found in the searches consist of various close-packed polytypes. Any repeating sequence of hexagonal layers, where no two adjacent layers are of the same type (the layer types are labeled A, B and C; see Fig.\ \ref{fig:hexagonal_layers}), gives a close-packed structure. As a result, there is an infinite number of different close packed polytypes. Many of these are realized in the Martensitic phase of Li that is observed experimentally at ambient pressure and low temperatures; this phase consists of a mixture of close-packed phases and remnants of the high-temperature BCC phase \cite{lithium_new_groundstate}. This is an example of one of the most common forms of Martensitic transformation (BCC $\rightarrow$ close-packed) \cite{pre-martensitic_anomalies}, in which anharmonic effects may play an important role \cite{landau_theory_bcc_9r}. Close packed crystal structures with a short repeating sequence of hexagonal layers are more likely to be realized in such a mixed phase, as several repeats are more likely to fit between defects. Layer-layer interactions also diminish rapidly with inter-layer separation \cite{ackland_close_packed}. This justifies the truncation of the infinite set to smaller sequences that are below a certain repeat length. In particular we investigate the energetics of all of the possible close-packed Li structures with a sequence of 15 or fewer layers. There are 10,922 such structures\footnote{This number is $\sim \frac{2}{3} 2^{14}$. In general the number of close-packed structures with $\leq n$ hexagonal layers is $\sim \frac{2}{3}2^n$. The factor of $2^n$ arises from the number of leaves in a tree such as Fig.\ \ref{fig:hexagonal_layers} and the $\frac{2}{3}$ arises from the $\frac{1}{3}$ of such structures which have the same initial and final layer in the sequence (meaning they do not correspond to a close packed structure).}, 489 of which are unique, labeling equivalent structures following 
Ref.\ \cite{ackland_close_packed}\footnote{We note a minor error in \cite{ackland_close_packed}, where it is stated that the number of unique close-packed sequences with up to 10 atomic layers is 43. There are in fact only 38.}. For a given unique structure $i$, we call the number of equivalent structures $w_i$.

\begin{figure*}
    \centering
    \includegraphics[width=\textwidth]{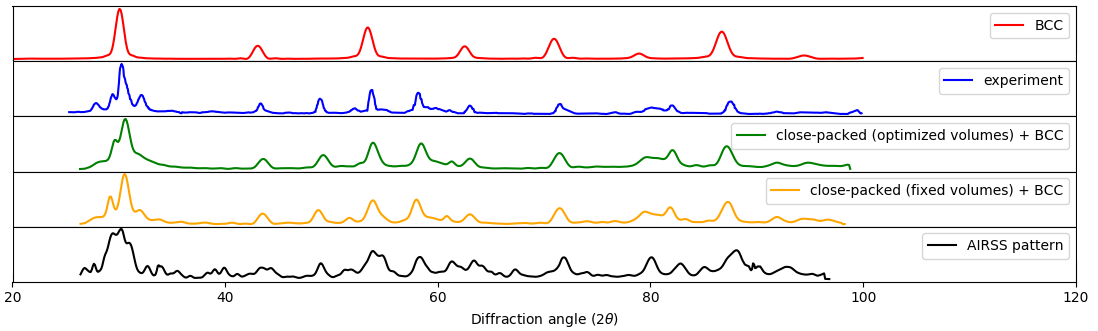}
    \caption{Simulated and experimental neutron diffraction patterns for Li. The experimental pattern ($<20$ K at ambient pressure) from \cite{mixed_phase_neutron_data} is shown, as well as patterns derived by thermodynamically weighting close-packed Li structures and AIRSS structures using DFT energies (see Eq.\ \ref{eq:neutron_pattern_weighting}). For the close-packed patterns all unique close-packed sequences consisting of fewer than 16 hexagonal layers are included. A Gaussian broadening was applied to the derived patterns with a width derived from a best-fit to the experimental data.}
    \label{fig:dft_powder}
\end{figure*}

The powder neutron diffraction pattern is simulated for each of these unique structures and the resulting overall mixed-phase pattern is obtained by thermodynamic weighting of the individual patterns. Neglecting defect energies the resulting neutron pattern at temperature $T$ is given by \cite{supplement}:
\begin{equation}
\label{eq:neutron_pattern_weighting}
    I(2\theta) = aI_{\text{BCC}}(2\theta) + b\sum_i I_i(2\theta) \frac{w_i}{\exp{\left(\frac{E_i - \mu}{k_B T}\right)} -1}
\end{equation}
where $I_i(2\theta)$ is the pattern for the $i^\text{th}$ structure, $E_i$ is the energy per atom and $w_i$ is the multiplicity of the structure (as defined in the previous paragraph). $a$, $b$, $\mu$ and $T$ are fitted to the experimental pattern. The resulting combined patterns are shown in Fig.\ \ref{fig:dft_powder}.

In agreement with Ackland \textit{et al.} the experimental data are well described by a Martensite consisting of a mixture of close-packed phases and BCC remnants \cite{lithium_new_groundstate}. We give the results for such weighted close-packed patterns both at a fixed volume per atom, and after optimizing the volume of each phase. The latter case is equivalent to allowing density variations across the sample based on the local crystal structure. We see little difference in the resulting patterns, likely because the optimized volumes are all very similar.

Simply weighting the close packed phases thermodynamically in this way reproduces many of the non-BCC characteristics of the experimental pattern, but not as accurately as the large-scale molecular dynamics calculations performed in Ref.\ \cite{lithium_new_groundstate}. This is because transformation kinetics are not fully included by a simple thermodynamic weighting of phases, as energy barriers along transition pathways between phases also play an important role \cite{minimum_energy_transition_paths}. Such kinetic effects can be captured by molecular dynamics simulations using suitable interatomic potentials \cite{lithium_new_groundstate, interatomic_potentials}. We are investigating the exploration of transition states from first principles in order to better understand these effects.

For completeness, the pattern that results from combining the structures found in our AIRSS search using Eq.\ \ref{eq:neutron_pattern_weighting} is also given in Fig.\ \ref{fig:dft_powder}.

\begin{table}
    \centering
    \begin{tabular}{|l|l|l|l|}
        \hline
        Space group & $F_{min}$ (meV/atom) & $V$ (\r{A}$^3$/atom) & $N$ \\
        \hline
$Fm\bar{3}m$ (FCC) & 0.0     & 18.96827 & 39 \\
$P6_3/mmc$ (HCP)   & 0.8115  & 18.97343 & 17 \\
$Im\bar{3}m$ (BCC) & 1.3684  & 18.94680 & 26 \\
$R\bar{3}m$ (9R)   & 2.0593  & 18.94582 & 3  \\
$I4/mmm$     & 2.1320  & 18.94480 & 30 \\
$C2/m$       & 2.5273  & 18.97458 & 10 \\
$Cmcm$       & 2.7300  & 18.93219 & 5  \\
$Immm$       & 5.2318  & 18.97877 & 2  \\
$P2_1/m$     & 11.781  & 18.99312 & 3  \\
$P6/mmm$     & 14.167  & 19.05003 & 3  \\
        \hline
    \end{tabular}
    \caption{The lowest-energy results of a simple Li AIRSS search. For each space group the number of times it was found ($N$) is shown. We also report, for each space group, the lowest Helmholtz free energy found ($F_{min}$, relative to the FCC phase) and the corresponding volume ($V$). The space groups are reported in order of increasing $F_{min}$. Only electronic energies are calculated, vibrational energies are neglected in these calculations. The conventional names for the four phases which are investigated more deeply in this work are shown in brackets next to the space group. In total 469 structures were generated \cite{supplement}. It is interesting to note the absence of the DHCP structure (see supplement \cite{supplement}).}
    \label{tab:AIRSS_ht}
\end{table}

\subsection{Free energy calculations}
\label{sec:quasi-harmonic_results}
\begin{figure}
    \includegraphics[width=\figurewidth]{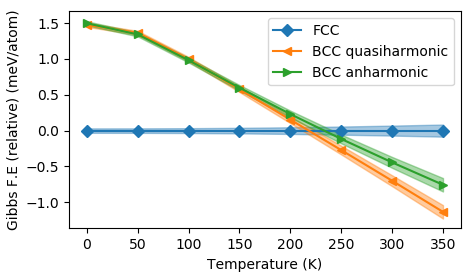}
    \caption{The Gibbs free energy of the FCC and BCC Li phases at a range of temperatures. This data was obtained by fitting the Birch-Murnaghan equation of state to data for $F(V,T)$ from DFT calculations (see main text). The standard fitting error is shown as a shaded region. The result of including the anharmonic correction from Fig.\ \ref{fig:anharmonic_correction} is also shown.}
    \label{fig:quasi-harmonic_ht}
\end{figure}

Quasi-harmonic calculations of the Gibbs free energy are performed for the FCC and BCC phases of Lithium, using a well-converged parameter set. We use plane-wave DFT with a PBE functional, a plane-wave cutoff of 3 keV and an electronic k-point grid with a spacing of 0.02 \r{A}$^{-1}$. The phonon Brillouin zone is sampled using the highly-efficient non-diagonal supercells method \cite{NONDIAGONAL_PHONON} which is significantly faster than a normal phonon supercell calculation, without any loss of accuracy. The phonon dispersion is calculated on a 10x10x10 grid in reciprocal space and then interpolated to a 40x40x40 grid using Fourier interpolation. The Helmholtz free energy is calculated for a range of volumes around the equilibrium volume and fitted to the Birch-Murnaghan equation of state \cite{birch_murnaghan} to extract the Gibbs free energy. An example of such a fit at 300 K is shown in Fig.\ \ref{fig:bm_example_300k}. The resulting Gibbs free energy is shown in Fig.\ \ref{fig:quasi-harmonic_ht} for a range of temperatures between 0 and 350 K. The Murnaghan, Rose-Vinet and Poirier-Tarantola equations of state \cite{murnaghan, rose-vinet, poirier-tarantola} give essentially identical results. A transition from the FCC to a BCC phase occurs upon heating at 217$\pm$13 K. This phase transition is observed on isobaric heating of the FCC phase experimentally; however, the calculated transition temperature is somewhat above the experimental range of 110-200 K 
\cite{lithium_new_groundstate,9r_cycling,9r_bcc_transition_smith, 9r_bcc_transition_anharmonicity, 9r_bcc_transition_schwarz, martensite_soft_xray}.  The reverse BCC $\rightarrow$ FCC transition is not seen experimentally upon isobaric cooling at ambient pressure. The FCC phase is instead prepared via a high-pressure route \cite{lithium_new_groundstate} to avoid formation of the Martensitic phase investigated in section \ref{sec:close_packed_structures}. Isothermal compressibilities and thermal expansion coefficients closely match experimental results \cite{supplement, kittel_textbook, li_thermal_expansion_coeff}.

\begin{figure}
    \includegraphics[width=\figurewidth]{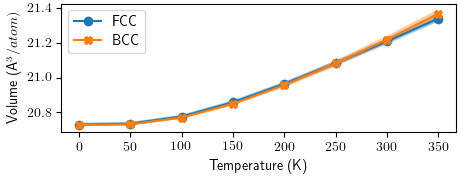}
    \caption{From the same calculation as Fig.\ \ref{fig:quasi-harmonic_ht}, but showing the derived equilibrium volume and thermal expansion.}
    \label{fig:quasi-harmonic_ht_volume}
\end{figure}

\begin{figure}
    \includegraphics[width=\figurewidth]{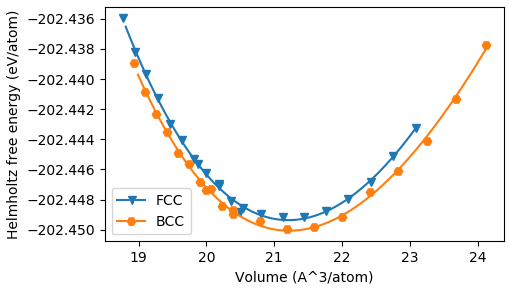}
    \caption{The Helmholtz free energy per atom verses volume per atom at 300 K for FCC and BCC Lithium. The solid lines show the Birch-Murnaghan fit used to derive the 300 K data in Figs.\ \ref{fig:quasi-harmonic_ht} and \ref{fig:quasi-harmonic_ht_volume}.}
    \label{fig:bm_example_300k}
\end{figure}

\begin{figure}
    \includegraphics[width=\figurewidth]{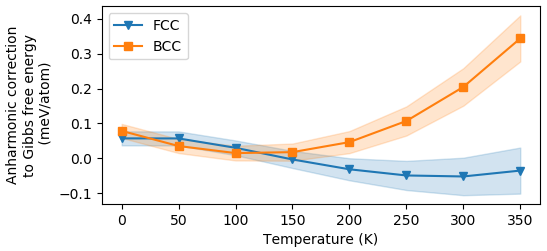}
    \caption{The correction to the Gibbs free energy of BCC and FCC Li that results from including the effects of anharmonic vibrations via Eq.\ \ref{eq:principle_axis_Hamiltonian}.}
    \label{fig:anharmonic_correction}
\end{figure}

Using the method outlined in section \ref{sec:vibrations_theory}, we calculate an anharmonic correction to the Helmholtz free energy, effectively moving from the quasi-harmonic regime into the ``quasi-anharmonic". The resulting correction to the Gibbs free energy is shown in Fig.\ \ref{fig:anharmonic_correction}. The effect on the FCC $\rightarrow$ BCC transition is included in Fig.\ \ref{fig:quasi-harmonic_ht}. Anharmonic effects are much stronger in the BCC phase, with the FCC $\rightarrow$ BCC transition temperature increasing by 15 K (to 232 K) as a result.

\begin{figure}
    \centering
    \includegraphics[width=\figurewidth]{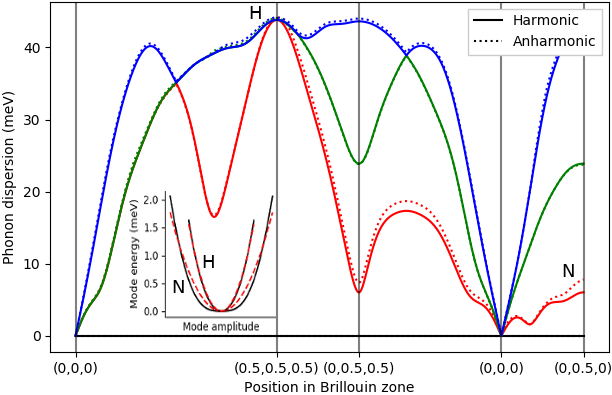}
    \caption{The 0\,K vibrational band structure of BCC Li. The solid line shows the normal harmonic phonon dispersion relation from Eq.\ \ref{eq:harmonic_hamiltonian}. The dotted line shows the lowest eigenvalue of the anharmonic Hamiltonian given in Eq.\ \ref{eq:principle_axis_Hamiltonian}. The Brillouin zone positions are given in terms of primitive reciprocal lattice vectors. The shapes of the mode potentials at points N and H in the Brillouin zone are shown in the inset with quadratic fits (red dashed lines) to illustrate the anharmonicity at the N point.}
    \label{fig:bcc_phonon_dispersion}
\end{figure}

\begin{figure}
    \centering
    \includegraphics[width=\figurewidth]{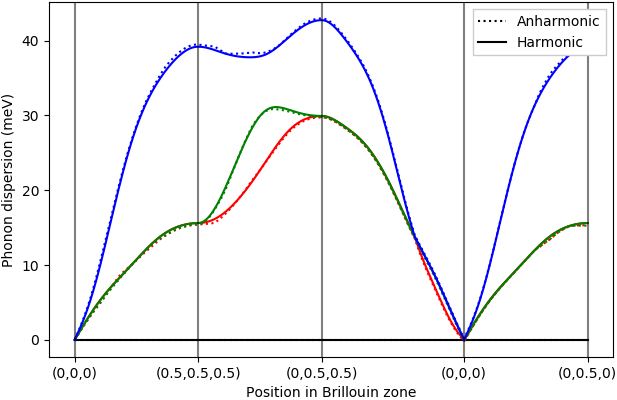}
    \caption{As Fig.\ \ref{fig:bcc_phonon_dispersion}, but for FCC Li. In this case the anharmonicity is negligible.}
    \label{fig:fcc_phonon_dispersion}
\end{figure}

It is also interesting to note the difference between the lowest-lying eigenvalues of the harmonic and anharmonic Hamiltonians (Eq.\ \ref{eq:harmonic_hamiltonian} and \ref{eq:principle_axis_Hamiltonian} respectively), shown in Figs. \ref{fig:bcc_phonon_dispersion} and \ref{fig:fcc_phonon_dispersion}. These corrections are present even at zero temperature as they result directly from the anharmonicity of the 0 K Born-Oppenheimer surface. The most significant effect is a stiffening of the N-point $(0, 0.5, 0)$ phonon in BCC Li by $\sim$ 1 meV (around 30 \% stiffer than the harmonic case). It has been suggested that stiffening of this mode could be important in stabilizing BCC phases \cite{n_point_anomalous_2}, however we find that this mode is dynamically stable to begin with (contrary to BCC Zr in \cite{n_point_anomalous_1, n_point_anomalous_2}).

\begin{figure}
    \includegraphics[width=\figurewidth]{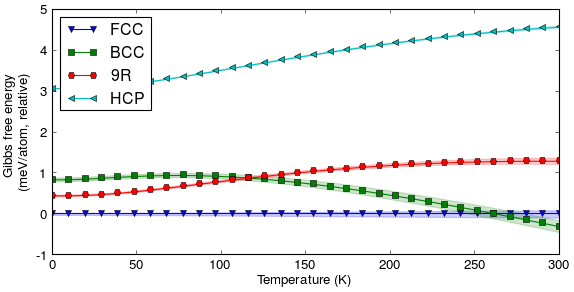}
    \caption{The Gibbs free energy of Li phases at a range of temperatures. The data was obtained by fitting the Birch-Murnaghan equation of state to data for $F(V,T)$, obtained using DFT calculations (see main text). The standard fitting error is shown as a shaded region.}
    \label{fig:quasi-harmonic_lt}
\end{figure}

To complement these results we perform a second set of similar calculations which include the 9R and HCP structures. These calculations are carried out with an LDA functional and with less intensive convergence parameters (1 keV plane-wave cutoff, 0.035 \r{A}$^{-1}$ electronic k-point grid spacing and an 8x8x8 phonon q-point grid). This is done in order to investigate the sensitivity of our results with respect to convergence parameters and the functional used. The resulting Gibbs free energies are shown in Fig.\ \ref{fig:quasi-harmonic_lt}. The sequence of phase transitions with increasing temperature is the same as in the PBE calculations in Fig.\ \ref{fig:quasi-harmonic_ht}. However, the FCC $\rightarrow$ BCC transition temperature in the LDA calculations is 47 K higher at 264$\pm$26 K. This is due to the fact that the transition temperature is extremely sensitive to the convergence of the differences in the Gibbs free energy; we estimate an increase of 117 K in transition temperature per meV increase in the BCC-FCC energy difference. The appearance of a metastable BCC $\rightarrow$ 9R transition upon cooling through 120 K is compatible with the Martensitic transition, however the predicted transition temperature is once again somewhat above the experimental range \cite{9r_bcc_transition_smith, 9r_bcc_transition_schwarz, pressure_effects_martensitic_transformation, 9r_bcc_transition_anharmonicity, 9r_cycling, martensite_soft_xray}.

We go on to show that the FCC $\rightarrow$ BCC transition is driven by entropic contributions to the phonon free energy. The phonon free energy is given by:
\begin{equation}
\begin{split}
    F_{ph}(T,V) = \frac{1}{2} &\int \omega D(\omega,V) d\omega \\ +&\int \frac{ \omega}{\exp{(\omega/T)}-1}D(\omega)d\omega - TS_{ph}(T,V)\\
\end{split}
\end{equation}
where $D(\omega)$ is the phonon density of states. The first term is the zero-point energy, the second arises from the thermal occupation of phonon states and the entropic term, $S_{ph}$, is given by \cite{phonon_entropy_theory,phonons_in_dft}:
\begin{equation}
\begin{split}
    S_{ph}(T,V) = &\int \frac{\omega/T}{\exp{(\omega/T)}-1}D(\omega, V)d\omega \\-&\int \ln{[1-\exp{(-\omega/T)}]}D(\omega,V)d\omega
\end{split}
\end{equation}

\begin{figure}
    \includegraphics[width=\figurewidth]{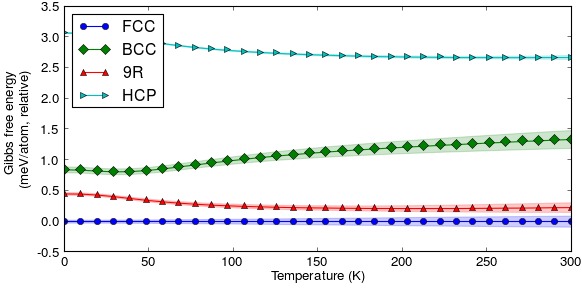}
    \caption{As Fig.\ \ref{fig:quasi-harmonic_lt}, but neglecting the entropic contribution to the phonon free energy. With this modification we see that the phase transitions in the 0--300 K range disappear.}
    \label{fig:quasi-harmonic_lt_noentropy}
\end{figure}

\begin{figure}
    \includegraphics[width=\figurewidth]{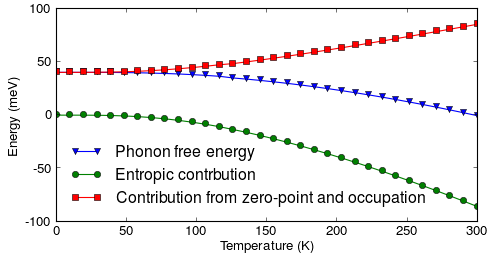}
    \caption{Phonon contributions to the Gibbs free energy of BCC Li.}
    \label{fig:bcc_li_phonon_fe_contrib}
\end{figure}

The importance of the entropic term becomes apparent when its contribution is explicitly neglected, resulting in the Gibbs free energy landscape shown in Fig.\ \ref{fig:quasi-harmonic_lt_noentropy} in which phase transitions are no longer present. The individual contributions to the phonon free energy for BCC Li are shown in Fig.\ \ref{fig:bcc_li_phonon_fe_contrib}, in which clearly that the entropic contribution dominates the zero-point and occupational terms. The FCC, HCP and 9R phases show similar behaviour. Because the low-lying phonon modes in BCC Li are softer than those of FCC (see Figs.\ \ref{fig:bcc_phonon_dispersion} and \ref{fig:fcc_phonon_dispersion}), the entropic effects are stronger, which leads to the FCC $\rightarrow$ BCC phase transition upon heating.

\section{Conclusions}   
We have investigated the application of a general framework for deriving the phase behaviour of materials with strong vibrations from first principles to metallic Li. The AIRSS approach \cite{AIRSS,airss_perspective} is found to quickly pick out relevant low-energy phases with little effort, including both the BCC phase and numerous close-packed phases. We have found that DFT predicts a zero-temperature FCC phase and a room temperature BCC phase; a metastable transition from BCC to the 9R phase is also predicted in the region of FCC stability. These results are consistent with experimental results, which show a (partial) Martensitic transition from BCC to various close-packed forms on cooling \cite{lithium_new_groundstate,9r_bcc_transition_smith, 9r_bcc_transition_anharmonicity, 9r_bcc_transition_schwarz}. Experiments also show that the FCC phase is stable at low temperatures, but may be formed via a high-pressure pathway \cite{lithium_new_groundstate}. The predicted FCC $\rightarrow$ BCC transition temperature is found to be extremely sensitive to changes in the Gibbs free energy landscape ($\sim$ 117 K/meV). As a result, effects which would normally be considered negligible become potentially important, including the effects of anharmonic vibrations. We calculate an anharmonic correction of up to 0.4 meV/atom, which increases the predicted FCC $\rightarrow$ BCC transition temperature by 15 K to 232 K, but find that it does not qualitatively change the phase diagram. Our calculations show that the N-point phonon in BCC Li is dynamically stable, contrary to the case for BCC Zr in Ref.\ \cite{n_point_anomalous_1,n_point_anomalous_2} and is around 30\% stiffer when anharmonic effects are included. The anharmonic free energy calculations are around 50 times more expensive than the quasi-harmonic ones, but we have found that they can be accelerated without loss of accuracy by employing symmetry considerations and sampling anharmonic potentials at fewer amplitudes. We calculate the various contributions to the Gibbs free energy and find that entropic contributions to the phonon free energy are of crucial importance, without which such phase transitions disappear entirely.

\begin{acknowledgements}
M.H.\ would like to acknowledge the EPSRC Centre for Doctoral Training in Computational Methods for Materials Science for funding under grant number EP/L015552/1. We thank Bartomeu Monserrat, Joseph Nelson and Nick Woods for helpful discussions. R.\ J.\ Needs acknowledges financial support from EPSRC under Critical Mass Grant EP/P034616/1 and the Support for the UKCP consortium grant EP/P022596/1. This work was performed using resources provided by the Cambridge Service for Data Driven Discovery (CSD3) operated by the University of Cambridge Research Computing Service (\weblink{http://www.csd3.cam.ac.uk/}), provided by Dell EMC and Intel using Tier-2 funding from the Engineering and Physical Sciences Research Council (capital grant EP/P020259/1), and DiRAC funding from the Science and Technology Facilities Council (\weblink{www.dirac.ac.uk}). Supporting research data may be freely accessed at \weblink{https://doi.org/10.17863/CAM.31655}, in compliance with the applicable Open Data policies.
\end{acknowledgements}

\bibliography{references.bib}

\begin{thebibliography}{42}%
\makeatletter
\providecommand \@ifxundefined [1]{%
 \@ifx{#1\undefined}
}%
\providecommand \@ifnum [1]{%
 \ifnum #1\expandafter \@firstoftwo
 \else \expandafter \@secondoftwo
 \fi
}%
\providecommand \@ifx [1]{%
 \ifx #1\expandafter \@firstoftwo
 \else \expandafter \@secondoftwo
 \fi
}%
\providecommand \natexlab [1]{#1}%
\providecommand \enquote  [1]{``#1''}%
\providecommand \bibnamefont  [1]{#1}%
\providecommand \bibfnamefont [1]{#1}%
\providecommand \citenamefont [1]{#1}%
\providecommand \href@noop [0]{\@secondoftwo}%
\providecommand \href [0]{\begingroup \@sanitize@url \@href}%
\providecommand \@href[1]{\@@startlink{#1}\@@href}%
\providecommand \@@href[1]{\endgroup#1\@@endlink}%
\providecommand \@sanitize@url [0]{\catcode `\\12\catcode `\$12\catcode
  `\&12\catcode `\#12\catcode `\^12\catcode `\_12\catcode `\%12\relax}%
\providecommand \@@startlink[1]{}%
\providecommand \@@endlink[0]{}%
\providecommand \url  [0]{\begingroup\@sanitize@url \@url }%
\providecommand \@url [1]{\endgroup\@href {#1}{\urlprefix }}%
\providecommand \urlprefix  [0]{URL }%
\providecommand \Eprint [0]{\href }%
\providecommand \doibase [0]{http://dx.doi.org/}%
\providecommand \selectlanguage [0]{\@gobble}%
\providecommand \bibinfo  [0]{\@secondoftwo}%
\providecommand \bibfield  [0]{\@secondoftwo}%
\providecommand \translation [1]{[#1]}%
\providecommand \BibitemOpen [0]{}%
\providecommand \bibitemStop [0]{}%
\providecommand \bibitemNoStop [0]{.\EOS\space}%
\providecommand \EOS [0]{\spacefactor3000\relax}%
\providecommand \BibitemShut  [1]{\csname bibitem#1\endcsname}%
\let\auto@bib@innerbib\@empty
\bibitem [{\citenamefont {Ackland}\ \emph {et~al.}(2017)\citenamefont
  {Ackland}, \citenamefont {Dunuwille}, \citenamefont {Martinez-Canales},
  \citenamefont {Loa}, \citenamefont {Zhang}, \citenamefont {Sinogeikin},
  \citenamefont {Cai},\ and\ \citenamefont
  {Deemyad}}]{lithium_new_groundstate}%
  \BibitemOpen
  \bibfield  {author} {\bibinfo {author} {\bibfnamefont {Graeme~J.}\
  \bibnamefont {Ackland}}, \bibinfo {author} {\bibfnamefont {Mihindra}\
  \bibnamefont {Dunuwille}}, \bibinfo {author} {\bibfnamefont {Miguel}\
  \bibnamefont {Martinez-Canales}}, \bibinfo {author} {\bibfnamefont {Ingo}\
  \bibnamefont {Loa}}, \bibinfo {author} {\bibfnamefont {Rong}\ \bibnamefont
  {Zhang}}, \bibinfo {author} {\bibfnamefont {Stanislav}\ \bibnamefont
  {Sinogeikin}}, \bibinfo {author} {\bibfnamefont {Weizhao}\ \bibnamefont
  {Cai}}, \ and\ \bibinfo {author} {\bibfnamefont {Shanti}\ \bibnamefont
  {Deemyad}},\ }\bibfield  {title} {\enquote {\bibinfo {title} {Quantum and
  isotope effects in lithium metal},}\ }\href {\doibase
  10.1126/science.aal4886} {\bibfield  {journal} {\bibinfo  {journal}
  {Science}\ }\textbf {\bibinfo {volume} {356}},\ \bibinfo {pages} {1254--1259}
  (\bibinfo {year} {2017})}\BibitemShut {NoStop}%
\bibitem [{\citenamefont {Schaeffer}\ \emph {et~al.}(2015)\citenamefont
  {Schaeffer}, \citenamefont {Cai}, \citenamefont {Olejnik}, \citenamefont
  {Molaison}, \citenamefont {Sinogeikin}, \citenamefont {dos Santos},\ and\
  \citenamefont {Deemyad}}]{li7_martensite}%
  \BibitemOpen
  \bibfield  {author} {\bibinfo {author} {\bibfnamefont {Anne~Marie}\
  \bibnamefont {Schaeffer}}, \bibinfo {author} {\bibfnamefont {Weizhao}\
  \bibnamefont {Cai}}, \bibinfo {author} {\bibfnamefont {Ella}\ \bibnamefont
  {Olejnik}}, \bibinfo {author} {\bibfnamefont {Jamie~J.}\ \bibnamefont
  {Molaison}}, \bibinfo {author} {\bibfnamefont {Stanislav}\ \bibnamefont
  {Sinogeikin}}, \bibinfo {author} {\bibfnamefont {Antonio~M.}\ \bibnamefont
  {dos Santos}}, \ and\ \bibinfo {author} {\bibfnamefont {Shanti}\ \bibnamefont
  {Deemyad}},\ }\bibfield  {title} {\enquote {\bibinfo {title} {Boundaries for
  martensitic transition of $^7$\text{Li} under pressure},}\ }\href {\doibase
  10.1038/ncomms9030} {\bibfield  {journal} {\bibinfo  {journal} {Nature
  Communications}\ }\textbf {\bibinfo {volume} {6}},\ \bibinfo {pages} {8030}
  (\bibinfo {year} {2015})}\BibitemShut {NoStop}%
\bibitem [{\citenamefont {Deemyad}\ and\ \citenamefont
  {Zhang}(2018)}]{quantum_effects_in_lithium}%
  \BibitemOpen
  \bibfield  {author} {\bibinfo {author} {\bibfnamefont {Shanti}\ \bibnamefont
  {Deemyad}}\ and\ \bibinfo {author} {\bibfnamefont {Rong}\ \bibnamefont
  {Zhang}},\ }\bibfield  {title} {\enquote {\bibinfo {title} {Probing quantum
  effects in lithium},}\ }\href {\doibase 10.1016/j.physc.2018.02.007}
  {\bibfield  {journal} {\bibinfo  {journal} {Physica C: Superconductivity and
  its Applications}\ }\textbf {\bibinfo {volume} {548}},\ \bibinfo {pages} {68
  -- 71} (\bibinfo {year} {2018})}\BibitemShut {NoStop}%
\bibitem [{\citenamefont {Pichl}\ \emph {et~al.}(2003)\citenamefont {Pichl},
  \citenamefont {Krystian}, \citenamefont {Prem},\ and\ \citenamefont
  {Krexner}}]{9r_cycling}%
  \BibitemOpen
  \bibfield  {author} {\bibinfo {author} {\bibfnamefont {W.}~\bibnamefont
  {Pichl}}, \bibinfo {author} {\bibfnamefont {M.}~\bibnamefont {Krystian}},
  \bibinfo {author} {\bibfnamefont {M.}~\bibnamefont {Prem}}, \ and\ \bibinfo
  {author} {\bibfnamefont {G.}~\bibnamefont {Krexner}},\ }\bibfield  {title}
  {\enquote {\bibinfo {title} {The martensite phase of high-purity lithium},}\
  }\href {\doibase 10.1051/jp4:20031073} {\bibfield  {journal} {\bibinfo
  {journal} {J. Phys. IV France}\ }\textbf {\bibinfo {volume} {112}},\ \bibinfo
  {pages} {1095--1098} (\bibinfo {year} {2003})}\BibitemShut {NoStop}%
\bibitem [{\citenamefont {Dmitrieva}\ \emph {et~al.}(2011)\citenamefont
  {Dmitrieva}, \citenamefont {Ponge}, \citenamefont {Inden}, \citenamefont
  {Millán}, \citenamefont {Choi}, \citenamefont {Sietsma},\ and\ \citenamefont
  {Raabe}}]{steel_martensite}%
  \BibitemOpen
  \bibfield  {author} {\bibinfo {author} {\bibfnamefont {O.}~\bibnamefont
  {Dmitrieva}}, \bibinfo {author} {\bibfnamefont {D.}~\bibnamefont {Ponge}},
  \bibinfo {author} {\bibfnamefont {G.}~\bibnamefont {Inden}}, \bibinfo
  {author} {\bibfnamefont {J.}~\bibnamefont {Millán}}, \bibinfo {author}
  {\bibfnamefont {P.}~\bibnamefont {Choi}}, \bibinfo {author} {\bibfnamefont
  {J.}~\bibnamefont {Sietsma}}, \ and\ \bibinfo {author} {\bibfnamefont
  {D.}~\bibnamefont {Raabe}},\ }\bibfield  {title} {\enquote {\bibinfo {title}
  {Chemical gradients across phase boundaries between martensite and austenite
  in steel studied by atom probe tomography and simulation},}\ }\href {\doibase
  10.1016/j.actamat.2010.09.042} {\bibfield  {journal} {\bibinfo  {journal}
  {Acta Materialia}\ }\textbf {\bibinfo {volume} {59}},\ \bibinfo {pages} {364
  -- 374} (\bibinfo {year} {2011})}\BibitemShut {NoStop}%
\bibitem [{\citenamefont {James}\ and\ \citenamefont
  {Hane}(2000)}]{shape-memory_materials}%
  \BibitemOpen
  \bibfield  {author} {\bibinfo {author} {\bibfnamefont {R.~D.}\ \bibnamefont
  {James}}\ and\ \bibinfo {author} {\bibfnamefont {K.~F.}\ \bibnamefont
  {Hane}},\ }\bibfield  {title} {\enquote {\bibinfo {title} {Martensitic
  transformations and shape-memory materials},}\ }\href {\doibase
  10.1016/S1359-6454(99)00295-5} {\bibfield  {journal} {\bibinfo  {journal}
  {Acta Materialia}\ }\textbf {\bibinfo {volume} {48}},\ \bibinfo {pages} {197
  -- 222} (\bibinfo {year} {2000})}\BibitemShut {NoStop}%
\bibitem [{\citenamefont {Ashcroft}(1989)}]{ashcroft_quantum_solid}%
  \BibitemOpen
  \bibfield  {author} {\bibinfo {author} {\bibfnamefont {N.~W.}\ \bibnamefont
  {Ashcroft}},\ }\bibfield  {title} {\enquote {\bibinfo {title} {Quantum-solid
  behavior and the electronic structure of the light alkali metals},}\ }\href
  {\doibase 10.1103/PhysRevB.39.10552} {\bibfield  {journal} {\bibinfo
  {journal} {Phys. Rev. B}\ }\textbf {\bibinfo {volume} {39}},\ \bibinfo
  {pages} {10552--10559} (\bibinfo {year} {1989})}\BibitemShut {NoStop}%
\bibitem [{\citenamefont {Staikov}\ \emph {et~al.}(1997)\citenamefont
  {Staikov}, \citenamefont {Kara},\ and\ \citenamefont
  {Rahman}}]{bulk_li_thermodynamics}%
  \BibitemOpen
  \bibfield  {author} {\bibinfo {author} {\bibfnamefont {P.}~\bibnamefont
  {Staikov}}, \bibinfo {author} {\bibfnamefont {A.}~\bibnamefont {Kara}}, \
  and\ \bibinfo {author} {\bibfnamefont {T.~S.}\ \bibnamefont {Rahman}},\
  }\bibfield  {title} {\enquote {\bibinfo {title} {First-principles studies of
  the thermodynamic properties of bulk \text{Li}},}\ }\href
  {http://stacks.iop.org/0953-8984/9/i=10/a=004} {\bibfield  {journal}
  {\bibinfo  {journal} {Journal of Physics: Condensed Matter}\ }\textbf
  {\bibinfo {volume} {9}},\ \bibinfo {pages} {2135} (\bibinfo {year}
  {1997})}\BibitemShut {NoStop}%
\bibitem [{\citenamefont {Caspersen}\ and\ \citenamefont
  {Carter}(2005)}]{minimum_energy_transition_paths}%
  \BibitemOpen
  \bibfield  {author} {\bibinfo {author} {\bibfnamefont {Kyle~J.}\ \bibnamefont
  {Caspersen}}\ and\ \bibinfo {author} {\bibfnamefont {Emily~A.}\ \bibnamefont
  {Carter}},\ }\bibfield  {title} {\enquote {\bibinfo {title} {Finding
  transition states for crystalline solid{\textendash}solid phase
  transformations},}\ }\href {\doibase 10.1073/pnas.0408127102} {\bibfield
  {journal} {\bibinfo  {journal} {Proceedings of the National Academy of
  Sciences}\ }\textbf {\bibinfo {volume} {102}},\ \bibinfo {pages} {6738--6743}
  (\bibinfo {year} {2005})}\BibitemShut {NoStop}%
\bibitem [{\citenamefont {Ackland}\ and\ \citenamefont
  {Jones}(2006)}]{local_crystal_structure}%
  \BibitemOpen
  \bibfield  {author} {\bibinfo {author} {\bibfnamefont {G.~J.}\ \bibnamefont
  {Ackland}}\ and\ \bibinfo {author} {\bibfnamefont {A.~P.}\ \bibnamefont
  {Jones}},\ }\bibfield  {title} {\enquote {\bibinfo {title} {Applications of
  local crystal structure measures in experiment and simulation},}\ }\href
  {\doibase 10.1103/PhysRevB.73.054104} {\bibfield  {journal} {\bibinfo
  {journal} {Phys. Rev. B}\ }\textbf {\bibinfo {volume} {73}},\ \bibinfo
  {pages} {054104} (\bibinfo {year} {2006})}\BibitemShut {NoStop}%
\bibitem [{\citenamefont {Naumov}\ \emph {et~al.}(2015)\citenamefont {Naumov},
  \citenamefont {Hemley}, \citenamefont {Hoffmann},\ and\ \citenamefont
  {Ashcroft}}]{li_free_electron_like}%
  \BibitemOpen
  \bibfield  {author} {\bibinfo {author} {\bibfnamefont {Ivan~I.}\ \bibnamefont
  {Naumov}}, \bibinfo {author} {\bibfnamefont {Russell~J.}\ \bibnamefont
  {Hemley}}, \bibinfo {author} {\bibfnamefont {Roald}\ \bibnamefont
  {Hoffmann}}, \ and\ \bibinfo {author} {\bibfnamefont {N.~W.}\ \bibnamefont
  {Ashcroft}},\ }\bibfield  {title} {\enquote {\bibinfo {title} {Chemical
  bonding in hydrogen and lithium under pressure},}\ }\href {\doibase
  10.1063/1.4928076} {\bibfield  {journal} {\bibinfo  {journal} {The Journal of
  Chemical Physics}\ }\textbf {\bibinfo {volume} {143}},\ \bibinfo {pages}
  {064702} (\bibinfo {year} {2015})}\BibitemShut {NoStop}%
\bibitem [{\citenamefont {Clark}\ \emph {et~al.}(2005)\citenamefont {Clark},
  \citenamefont {Segall}, \citenamefont {Pickard}, \citenamefont {Hasnip},
  \citenamefont {Probert}, \citenamefont {Refson},\ and\ \citenamefont
  {Payne}}]{castep}%
  \BibitemOpen
  \bibfield  {author} {\bibinfo {author} {\bibfnamefont {S.~J.}\ \bibnamefont
  {Clark}}, \bibinfo {author} {\bibfnamefont {M.~D.}\ \bibnamefont {Segall}},
  \bibinfo {author} {\bibfnamefont {C.~J.}\ \bibnamefont {Pickard}}, \bibinfo
  {author} {\bibfnamefont {P.~J.}\ \bibnamefont {Hasnip}}, \bibinfo {author}
  {\bibfnamefont {M.~J.}\ \bibnamefont {Probert}}, \bibinfo {author}
  {\bibfnamefont {K.}~\bibnamefont {Refson}}, \ and\ \bibinfo {author}
  {\bibfnamefont {M.~C.}\ \bibnamefont {Payne}},\ }\bibfield  {title} {\enquote
  {\bibinfo {title} {First principles methods using \textsc{CASTEP}},}\ }\href
  {http://eprints.whiterose.ac.uk/8521/} {\bibfield  {journal} {\bibinfo
  {journal} {Zeitschrift f{\"u}r Kristallographie}\ ,\ \bibinfo {pages}
  {567--570}} (\bibinfo {year} {2005})}\BibitemShut {NoStop}%
\bibitem [{\citenamefont {Pickard}\ and\ \citenamefont {Needs}(2011)}]{AIRSS}%
  \BibitemOpen
  \bibfield  {author} {\bibinfo {author} {\bibfnamefont {Chris~J.}\
  \bibnamefont {Pickard}}\ and\ \bibinfo {author} {\bibfnamefont {R.~J.}\
  \bibnamefont {Needs}},\ }\bibfield  {title} {\enquote {\bibinfo {title} {Ab
  initio random structure searching},}\ }\href {\doibase
  10.1088/0953-8984/23/5/053201} {\bibfield  {journal} {\bibinfo  {journal} {J.
  Phys.: Condens. Matter}\ }\textbf {\bibinfo {volume} {23}},\ \bibinfo {pages}
  {053201} (\bibinfo {year} {2011})}\BibitemShut {NoStop}%
\bibitem [{\citenamefont {Needs}\ and\ \citenamefont
  {Pickard}(2016)}]{airss_perspective}%
  \BibitemOpen
  \bibfield  {author} {\bibinfo {author} {\bibfnamefont {Richard~J.}\
  \bibnamefont {Needs}}\ and\ \bibinfo {author} {\bibfnamefont {Chris~J.}\
  \bibnamefont {Pickard}},\ }\bibfield  {title} {\enquote {\bibinfo {title}
  {Perspective: Role of structure prediction in materials discovery and
  design},}\ }\href {\doibase 10.1063/1.4949361} {\bibfield  {journal}
  {\bibinfo  {journal} {APL Materials}\ }\textbf {\bibinfo {volume} {4}},\
  \bibinfo {pages} {053210} (\bibinfo {year} {2016})}\BibitemShut {NoStop}%
\bibitem [{\citenamefont {Pickard}\ and\ \citenamefont
  {Needs}(2006)}]{airss_silane}%
  \BibitemOpen
  \bibfield  {author} {\bibinfo {author} {\bibfnamefont {Chris~J.}\
  \bibnamefont {Pickard}}\ and\ \bibinfo {author} {\bibfnamefont {R.~J.}\
  \bibnamefont {Needs}},\ }\bibfield  {title} {\enquote {\bibinfo {title}
  {High-pressure phases of silane},}\ }\href {\doibase
  10.1103/PhysRevLett.97.045504} {\bibfield  {journal} {\bibinfo  {journal}
  {Phys. Rev. Lett.}\ }\textbf {\bibinfo {volume} {97}},\ \bibinfo {pages}
  {045504} (\bibinfo {year} {2006})}\BibitemShut {NoStop}%
\bibitem [{\citenamefont {Gorelli}\ \emph {et~al.}(2012)\citenamefont
  {Gorelli}, \citenamefont {Elatresh}, \citenamefont {Guillaume}, \citenamefont
  {Marqu\'es}, \citenamefont {Ackland}, \citenamefont {Santoro}, \citenamefont
  {Bonev},\ and\ \citenamefont {Gregoryanz}}]{latt_dyn_dense_li}%
  \BibitemOpen
  \bibfield  {author} {\bibinfo {author} {\bibfnamefont {F.~A.}\ \bibnamefont
  {Gorelli}}, \bibinfo {author} {\bibfnamefont {S.~F.}\ \bibnamefont
  {Elatresh}}, \bibinfo {author} {\bibfnamefont {C.~L.}\ \bibnamefont
  {Guillaume}}, \bibinfo {author} {\bibfnamefont {M.}~\bibnamefont
  {Marqu\'es}}, \bibinfo {author} {\bibfnamefont {G.~J.}\ \bibnamefont
  {Ackland}}, \bibinfo {author} {\bibfnamefont {M.}~\bibnamefont {Santoro}},
  \bibinfo {author} {\bibfnamefont {S.~A.}\ \bibnamefont {Bonev}}, \ and\
  \bibinfo {author} {\bibfnamefont {E.}~\bibnamefont {Gregoryanz}},\ }\bibfield
   {title} {\enquote {\bibinfo {title} {Lattice dynamics of dense lithium},}\
  }\href {\doibase 10.1103/PhysRevLett.108.055501} {\bibfield  {journal}
  {\bibinfo  {journal} {Phys. Rev. Lett.}\ }\textbf {\bibinfo {volume} {108}},\
  \bibinfo {pages} {055501} (\bibinfo {year} {2012})}\BibitemShut {NoStop}%
\bibitem [{\citenamefont {Monserrat}\ \emph {et~al.}(2013)\citenamefont
  {Monserrat}, \citenamefont {Drummond},\ and\ \citenamefont
  {Needs}}]{anharmonic_phonons}%
  \BibitemOpen
  \bibfield  {author} {\bibinfo {author} {\bibfnamefont {Bartomeu}\
  \bibnamefont {Monserrat}}, \bibinfo {author} {\bibfnamefont {N.~D.}\
  \bibnamefont {Drummond}}, \ and\ \bibinfo {author} {\bibfnamefont {R.~J.}\
  \bibnamefont {Needs}},\ }\bibfield  {title} {\enquote {\bibinfo {title}
  {Anharmonic vibrational properties in periodic systems: energy,
  electron-phonon coupling, and stress},}\ }\href {\doibase
  10.1103/PhysRevB.87.144302} {\bibfield  {journal} {\bibinfo  {journal} {Phys.
  Rev. B}\ }\textbf {\bibinfo {volume} {87}},\ \bibinfo {pages} {144302}
  (\bibinfo {year} {2013})}\BibitemShut {NoStop}%
\bibitem [{sup()}]{supplement}%
  \BibitemOpen
  \bibfield  {title} {\enquote {\bibinfo {title} {Supplementary information},}\
  }\href@noop {} {\ }\BibitemShut {NoStop}%
\bibitem [{\citenamefont {Fultz}(2010)}]{phonon_entropy_theory}%
  \BibitemOpen
  \bibfield  {author} {\bibinfo {author} {\bibfnamefont {Brent}\ \bibnamefont
  {Fultz}},\ }\bibfield  {title} {\enquote {\bibinfo {title} {Vibrational
  thermodynamics of materials},}\ }\href {\doibase
  10.1016/j.pmatsci.2009.05.002} {\bibfield  {journal} {\bibinfo  {journal}
  {Progress in Materials Science}\ }\textbf {\bibinfo {volume} {55}},\ \bibinfo
  {pages} {247 -- 352} (\bibinfo {year} {2010})}\BibitemShut {NoStop}%
\bibitem [{\citenamefont {Baroni}\ \emph {et~al.}(2001)\citenamefont {Baroni},
  \citenamefont {de~Gironcoli}, \citenamefont {Dal~Corso},\ and\ \citenamefont
  {Giannozzi}}]{phonons_in_dft}%
  \BibitemOpen
  \bibfield  {author} {\bibinfo {author} {\bibfnamefont {Stefano}\ \bibnamefont
  {Baroni}}, \bibinfo {author} {\bibfnamefont {Stefano}\ \bibnamefont
  {de~Gironcoli}}, \bibinfo {author} {\bibfnamefont {Andrea}\ \bibnamefont
  {Dal~Corso}}, \ and\ \bibinfo {author} {\bibfnamefont {Paolo}\ \bibnamefont
  {Giannozzi}},\ }\bibfield  {title} {\enquote {\bibinfo {title} {Phonons and
  related crystal properties from density-functional perturbation theory},}\
  }\href {\doibase 10.1103/RevModPhys.73.515} {\bibfield  {journal} {\bibinfo
  {journal} {Rev. Mod. Phys.}\ }\textbf {\bibinfo {volume} {73}},\ \bibinfo
  {pages} {515--562} (\bibinfo {year} {2001})}\BibitemShut {NoStop}%
\bibitem [{\citenamefont {Overhauser}(1984)}]{overhauser_Li}%
  \BibitemOpen
  \bibfield  {author} {\bibinfo {author} {\bibfnamefont {A.~W.}\ \bibnamefont
  {Overhauser}},\ }\bibfield  {title} {\enquote {\bibinfo {title} {Crystal
  structure of lithium at 4.2 \text{K}},}\ }\href {\doibase
  10.1103/PhysRevLett.53.64} {\bibfield  {journal} {\bibinfo  {journal} {Phys.
  Rev. Lett.}\ }\textbf {\bibinfo {volume} {53}},\ \bibinfo {pages} {64--65}
  (\bibinfo {year} {1984})}\BibitemShut {NoStop}%
\bibitem [{\citenamefont {Barrett}(1956)}]{alkali_metals_low_temp_xray}%
  \BibitemOpen
  \bibfield  {author} {\bibinfo {author} {\bibfnamefont {C.~S.}\ \bibnamefont
  {Barrett}},\ }\bibfield  {title} {\enquote {\bibinfo {title} {X-ray study of
  the alkali metals at low temperatures},}\ }\href {\doibase
  10.1107/S0365110X56001790} {\bibfield  {journal} {\bibinfo  {journal} {Acta
  Crystallographica}\ }\textbf {\bibinfo {volume} {9}},\ \bibinfo {pages}
  {671--677} (\bibinfo {year} {1956})}\BibitemShut {NoStop}%
\bibitem [{\citenamefont {Barrett}\ and\ \citenamefont
  {Trautz}(1948)}]{barrett_Li}%
  \BibitemOpen
  \bibfield  {author} {\bibinfo {author} {\bibfnamefont {C.~S.}\ \bibnamefont
  {Barrett}}\ and\ \bibinfo {author} {\bibfnamefont {O.~R.}\ \bibnamefont
  {Trautz}},\ }\bibfield  {title} {\enquote {\bibinfo {title} {Transactions},}\
  }\href@noop {} {\bibfield  {journal} {\bibinfo  {journal} {AIME}\ }\textbf
  {\bibinfo {volume} {175}},\ \bibinfo {pages} {579} (\bibinfo {year}
  {1948})}\BibitemShut {NoStop}%
\bibitem [{\citenamefont {Katsnelson}\ \emph {et~al.}(1994)\citenamefont
  {Katsnelson}, \citenamefont {Naumov},\ and\ \citenamefont
  {Trefilov}}]{pre-martensitic_anomalies}%
  \BibitemOpen
  \bibfield  {author} {\bibinfo {author} {\bibfnamefont {M.~I.}\ \bibnamefont
  {Katsnelson}}, \bibinfo {author} {\bibfnamefont {I.~I.}\ \bibnamefont
  {Naumov}}, \ and\ \bibinfo {author} {\bibfnamefont {A.~V.}\ \bibnamefont
  {Trefilov}},\ }\bibfield  {title} {\enquote {\bibinfo {title} {Singularities
  of the electronic structure and pre-martensitic anomalies of lattice
  properties in $\beta$-phases of metals and alloys},}\ }\href {\doibase
  10.1080/01411599408201172} {\bibfield  {journal} {\bibinfo  {journal} {Phase
  Transitions}\ }\textbf {\bibinfo {volume} {49}},\ \bibinfo {pages} {143--191}
  (\bibinfo {year} {1994})}\BibitemShut {NoStop}%
\bibitem [{\citenamefont {Gooding}\ and\ \citenamefont
  {Krumhansl}(1988)}]{landau_theory_bcc_9r}%
  \BibitemOpen
  \bibfield  {author} {\bibinfo {author} {\bibfnamefont {R.~J.}\ \bibnamefont
  {Gooding}}\ and\ \bibinfo {author} {\bibfnamefont {J.~A.}\ \bibnamefont
  {Krumhansl}},\ }\bibfield  {title} {\enquote {\bibinfo {title} {Theory of the
  \text{BCC}-to-\text{9R} structural phase transformation of \text{Li}},}\
  }\href {\doibase 10.1103/PhysRevB.38.1695} {\bibfield  {journal} {\bibinfo
  {journal} {Phys. Rev. B}\ }\textbf {\bibinfo {volume} {38}},\ \bibinfo
  {pages} {1695--1704} (\bibinfo {year} {1988})}\BibitemShut {NoStop}%
\bibitem [{\citenamefont {Loach}\ and\ \citenamefont
  {Ackland}(2017)}]{ackland_close_packed}%
  \BibitemOpen
  \bibfield  {author} {\bibinfo {author} {\bibfnamefont {Christian~H.}\
  \bibnamefont {Loach}}\ and\ \bibinfo {author} {\bibfnamefont {Graeme~J.}\
  \bibnamefont {Ackland}},\ }\bibfield  {title} {\enquote {\bibinfo {title}
  {Stacking characteristics of close packed materials},}\ }\href {\doibase
  10.1103/PhysRevLett.119.205701} {\bibfield  {journal} {\bibinfo  {journal}
  {Phys. Rev. Lett.}\ }\textbf {\bibinfo {volume} {119}},\ \bibinfo {pages}
  {205701} (\bibinfo {year} {2017})}\BibitemShut {NoStop}%
\bibitem [{\citenamefont {Berliner}\ and\ \citenamefont
  {Werner}(1986)}]{mixed_phase_neutron_data}%
  \BibitemOpen
  \bibfield  {author} {\bibinfo {author} {\bibfnamefont {R.}~\bibnamefont
  {Berliner}}\ and\ \bibinfo {author} {\bibfnamefont {S.~A.}\ \bibnamefont
  {Werner}},\ }\bibfield  {title} {\enquote {\bibinfo {title} {Effect of
  stacking faults on diffraction: The structure of lithium metal},}\ }\href
  {\doibase 10.1103/PhysRevB.34.3586} {\bibfield  {journal} {\bibinfo
  {journal} {Phys. Rev. B}\ }\textbf {\bibinfo {volume} {34}},\ \bibinfo
  {pages} {3586--3603} (\bibinfo {year} {1986})}\BibitemShut {NoStop}%
\bibitem [{\citenamefont {Ko}\ and\ \citenamefont
  {Jeon}(2017)}]{interatomic_potentials}%
  \BibitemOpen
  \bibfield  {author} {\bibinfo {author} {\bibfnamefont {Won-Seok}\
  \bibnamefont {Ko}}\ and\ \bibinfo {author} {\bibfnamefont {Jong~Bae}\
  \bibnamefont {Jeon}},\ }\bibfield  {title} {\enquote {\bibinfo {title}
  {Interatomic potential that describes martensitic phase transformations in
  pure lithium},}\ }\href {\doibase 10.1016/j.commatsci.2016.12.018} {\bibfield
   {journal} {\bibinfo  {journal} {Computational Materials Science}\ }\textbf
  {\bibinfo {volume} {129}},\ \bibinfo {pages} {202 -- 210} (\bibinfo {year}
  {2017})}\BibitemShut {NoStop}%
\bibitem [{\citenamefont {Lloyd-Williams}\ and\ \citenamefont
  {Monserrat}(2015)}]{NONDIAGONAL_PHONON}%
  \BibitemOpen
  \bibfield  {author} {\bibinfo {author} {\bibfnamefont {Jonathan~H.}\
  \bibnamefont {Lloyd-Williams}}\ and\ \bibinfo {author} {\bibfnamefont
  {Bartomeu}\ \bibnamefont {Monserrat}},\ }\bibfield  {title} {\enquote
  {\bibinfo {title} {Lattice dynamics and electron-phonon coupling using
  nondiagonal supercells},}\ }\href {\doibase 10.1103/PhysRevB.92.184301}
  {\bibfield  {journal} {\bibinfo  {journal} {Phys. Rev. B}\ }\textbf {\bibinfo
  {volume} {92}},\ \bibinfo {pages} {184301} (\bibinfo {year}
  {2015})}\BibitemShut {NoStop}%
\bibitem [{\citenamefont {Birch}(1947)}]{birch_murnaghan}%
  \BibitemOpen
  \bibfield  {author} {\bibinfo {author} {\bibfnamefont {Francis}\ \bibnamefont
  {Birch}},\ }\bibfield  {title} {\enquote {\bibinfo {title} {Finite elastic
  strain of cubic crystals},}\ }\href {\doibase 10.1103/PhysRev.71.809}
  {\bibfield  {journal} {\bibinfo  {journal} {Phys. Rev.}\ }\textbf {\bibinfo
  {volume} {71}},\ \bibinfo {pages} {809--824} (\bibinfo {year}
  {1947})}\BibitemShut {NoStop}%
\bibitem [{\citenamefont {Murnaghan}(1944)}]{murnaghan}%
  \BibitemOpen
  \bibfield  {author} {\bibinfo {author} {\bibfnamefont {F.~D.}\ \bibnamefont
  {Murnaghan}},\ }\bibfield  {title} {\enquote {\bibinfo {title} {The
  compressibility of media under extreme pressures},}\ }\href {\doibase
  10.1073/pnas.30.9.244} {\bibfield  {journal} {\bibinfo  {journal}
  {Proceedings of the National Academy of Sciences}\ }\textbf {\bibinfo
  {volume} {30}},\ \bibinfo {pages} {244--247} (\bibinfo {year} {1944})},\
  \Eprint {http://arxiv.org/abs/http://www.pnas.org/content/30/9/244.full.pdf}
  {http://www.pnas.org/content/30/9/244.full.pdf} \BibitemShut {NoStop}%
\bibitem [{\citenamefont {Vinet}\ \emph {et~al.}(1987)\citenamefont {Vinet},
  \citenamefont {Smith}, \citenamefont {Ferrante},\ and\ \citenamefont
  {Rose}}]{rose-vinet}%
  \BibitemOpen
  \bibfield  {author} {\bibinfo {author} {\bibfnamefont {Pascal}\ \bibnamefont
  {Vinet}}, \bibinfo {author} {\bibfnamefont {John~R.}\ \bibnamefont {Smith}},
  \bibinfo {author} {\bibfnamefont {John}\ \bibnamefont {Ferrante}}, \ and\
  \bibinfo {author} {\bibfnamefont {James~H.}\ \bibnamefont {Rose}},\
  }\bibfield  {title} {\enquote {\bibinfo {title} {Temperature effects on the
  universal equation of state of solids},}\ }\href {\doibase
  10.1103/PhysRevB.35.1945} {\bibfield  {journal} {\bibinfo  {journal} {Phys.
  Rev. B}\ }\textbf {\bibinfo {volume} {35}},\ \bibinfo {pages} {1945--1953}
  (\bibinfo {year} {1987})}\BibitemShut {NoStop}%
\bibitem [{\citenamefont {{Poirier}}\ and\ \citenamefont
  {{Tarantola}}(1998)}]{poirier-tarantola}%
  \BibitemOpen
  \bibfield  {author} {\bibinfo {author} {\bibfnamefont {J.-P.}\ \bibnamefont
  {{Poirier}}}\ and\ \bibinfo {author} {\bibfnamefont {A.}~\bibnamefont
  {{Tarantola}}},\ }\bibfield  {title} {\enquote {\bibinfo {title} {{A
  logarithmic equation of state}},}\ }\href {\doibase
  10.1016/S0031-9201(98)00112-5} {\bibfield  {journal} {\bibinfo  {journal}
  {Physics of the Earth and Planetary Interiors}\ }\textbf {\bibinfo {volume}
  {109}},\ \bibinfo {pages} {1--8} (\bibinfo {year} {1998})}\BibitemShut
  {NoStop}%
\bibitem [{\citenamefont {Smith}(1987)}]{9r_bcc_transition_smith}%
  \BibitemOpen
  \bibfield  {author} {\bibinfo {author} {\bibfnamefont {H.~G.}\ \bibnamefont
  {Smith}},\ }\bibfield  {title} {\enquote {\bibinfo {title} {Martensitic phase
  transformation of single-crystal lithium from bcc to a \text{9R}-related
  structure},}\ }\href {\doibase 10.1103/PhysRevLett.58.1228} {\bibfield
  {journal} {\bibinfo  {journal} {Phys. Rev. Lett.}\ }\textbf {\bibinfo
  {volume} {58}},\ \bibinfo {pages} {1228--1231} (\bibinfo {year}
  {1987})}\BibitemShut {NoStop}%
\bibitem [{\citenamefont {McCarthy}\ \emph {et~al.}(1980)\citenamefont
  {McCarthy}, \citenamefont {Tompson},\ and\ \citenamefont
  {Werner}}]{9r_bcc_transition_anharmonicity}%
  \BibitemOpen
  \bibfield  {author} {\bibinfo {author} {\bibfnamefont {C.~M.}\ \bibnamefont
  {McCarthy}}, \bibinfo {author} {\bibfnamefont {C.~W.}\ \bibnamefont
  {Tompson}}, \ and\ \bibinfo {author} {\bibfnamefont {S.~A.}\ \bibnamefont
  {Werner}},\ }\bibfield  {title} {\enquote {\bibinfo {title} {Anharmonicity
  and the low-temperature phase in lithium metal},}\ }\href {\doibase
  10.1103/PhysRevB.22.574} {\bibfield  {journal} {\bibinfo  {journal} {Phys.
  Rev. B}\ }\textbf {\bibinfo {volume} {22}},\ \bibinfo {pages} {574--580}
  (\bibinfo {year} {1980})}\BibitemShut {NoStop}%
\bibitem [{\citenamefont {Schwarz}\ and\ \citenamefont
  {Blaschko}(1990)}]{9r_bcc_transition_schwarz}%
  \BibitemOpen
  \bibfield  {author} {\bibinfo {author} {\bibfnamefont {W.}~\bibnamefont
  {Schwarz}}\ and\ \bibinfo {author} {\bibfnamefont {O.}~\bibnamefont
  {Blaschko}},\ }\bibfield  {title} {\enquote {\bibinfo {title} {Polytype
  structures of lithium at low temperatures},}\ }\href {\doibase
  10.1103/PhysRevLett.65.3144} {\bibfield  {journal} {\bibinfo  {journal}
  {Phys. Rev. Lett.}\ }\textbf {\bibinfo {volume} {65}},\ \bibinfo {pages}
  {3144--3147} (\bibinfo {year} {1990})}\BibitemShut {NoStop}%
\bibitem [{\citenamefont {Crisp}(1991)}]{martensite_soft_xray}%
  \BibitemOpen
  \bibfield  {author} {\bibinfo {author} {\bibfnamefont {R.~S.}\ \bibnamefont
  {Crisp}},\ }\bibfield  {title} {\enquote {\bibinfo {title} {Observation of
  the low-temperature martensitic transformation in \text{Li} and a
  \text{Li-Mg} alloy by soft x-ray emission},}\ }\href
  {http://stacks.iop.org/0953-8984/3/i=30/a=009} {\bibfield  {journal}
  {\bibinfo  {journal} {Journal of Physics: Condensed Matter}\ }\textbf
  {\bibinfo {volume} {3}},\ \bibinfo {pages} {5761} (\bibinfo {year}
  {1991})}\BibitemShut {NoStop}%
\bibitem [{\citenamefont {Kittel}(2005)}]{kittel_textbook}%
  \BibitemOpen
  \bibfield  {author} {\bibinfo {author} {\bibfnamefont {Charles}\ \bibnamefont
  {Kittel}},\ }\href@noop {} {\emph {\bibinfo {title} {Introduction to Solid
  State Physics, 8th edition.}}}\ (\bibinfo  {publisher} {Hoboken, NJ: John
  Wiley \& Sons, Inc},\ \bibinfo {year} {2005})\ p.~\bibinfo {pages}
  {50}\BibitemShut {NoStop}%
\bibitem [{\citenamefont {Cohen}\ \emph {et~al.}(2003)\citenamefont {Cohen},
  \citenamefont {Lide},\ and\ \citenamefont
  {Trigg}}]{li_thermal_expansion_coeff}%
  \BibitemOpen
  \bibfield  {author} {\bibinfo {author} {\bibfnamefont {E.~Richard}\
  \bibnamefont {Cohen}}, \bibinfo {author} {\bibfnamefont {David~R.}\
  \bibnamefont {Lide}}, \ and\ \bibinfo {author} {\bibfnamefont {George~L.}\
  \bibnamefont {Trigg}},\ }\href@noop {} {\emph {\bibinfo {title} {Physics Desk
  Reference, 3rd edition}}}\ (\bibinfo  {publisher} {Springer-Verlag New York,
  Inc.},\ \bibinfo {year} {2003})\ p.\ \bibinfo {pages} {826}\BibitemShut
  {NoStop}%
\bibitem [{\citenamefont {Chen}\ \emph {et~al.}(1985)\citenamefont {Chen},
  \citenamefont {Fu}, \citenamefont {Ho},\ and\ \citenamefont
  {Harmon}}]{n_point_anomalous_2}%
  \BibitemOpen
  \bibfield  {author} {\bibinfo {author} {\bibfnamefont {Y.}~\bibnamefont
  {Chen}}, \bibinfo {author} {\bibfnamefont {C.-L.}\ \bibnamefont {Fu}},
  \bibinfo {author} {\bibfnamefont {K.-M.}\ \bibnamefont {Ho}}, \ and\ \bibinfo
  {author} {\bibfnamefont {B.~N.}\ \bibnamefont {Harmon}},\ }\bibfield  {title}
  {\enquote {\bibinfo {title} {Calculations for the transverse n-point phonons
  in bcc zr, nb, and mo},}\ }\href {\doibase 10.1103/PhysRevB.31.6775}
  {\bibfield  {journal} {\bibinfo  {journal} {Phys. Rev. B}\ }\textbf {\bibinfo
  {volume} {31}},\ \bibinfo {pages} {6775--6778} (\bibinfo {year}
  {1985})}\BibitemShut {NoStop}%
\bibitem [{\citenamefont {Pinsook}\ and\ \citenamefont
  {Ackland}(1999)}]{n_point_anomalous_1}%
  \BibitemOpen
  \bibfield  {author} {\bibinfo {author} {\bibfnamefont {U.}~\bibnamefont
  {Pinsook}}\ and\ \bibinfo {author} {\bibfnamefont {G.~J.}\ \bibnamefont
  {Ackland}},\ }\bibfield  {title} {\enquote {\bibinfo {title} {Calculation of
  anomalous phonons and the hcp-bcc phase transition in zirconium},}\ }\href
  {\doibase 10.1103/PhysRevB.59.13642} {\bibfield  {journal} {\bibinfo
  {journal} {Phys. Rev. B}\ }\textbf {\bibinfo {volume} {59}},\ \bibinfo
  {pages} {13642--13649} (\bibinfo {year} {1999})}\BibitemShut {NoStop}%
\bibitem [{\citenamefont {Smith}\ \emph {et~al.}(1990)\citenamefont {Smith},
  \citenamefont {Berliner}, \citenamefont {Jorgensen}, \citenamefont
  {Nielsen},\ and\ \citenamefont
  {Trivisonno}}]{pressure_effects_martensitic_transformation}%
  \BibitemOpen
  \bibfield  {author} {\bibinfo {author} {\bibfnamefont {H.~G.}\ \bibnamefont
  {Smith}}, \bibinfo {author} {\bibfnamefont {R.}~\bibnamefont {Berliner}},
  \bibinfo {author} {\bibfnamefont {J.~D.}\ \bibnamefont {Jorgensen}}, \bibinfo
  {author} {\bibfnamefont {M.}~\bibnamefont {Nielsen}}, \ and\ \bibinfo
  {author} {\bibfnamefont {J.}~\bibnamefont {Trivisonno}},\ }\bibfield  {title}
  {\enquote {\bibinfo {title} {Pressure effects on the martensitic
  transformation in metallic lithium},}\ }\href {\doibase
  10.1103/PhysRevB.41.1231} {\bibfield  {journal} {\bibinfo  {journal} {Phys.
  Rev. B}\ }\textbf {\bibinfo {volume} {41}},\ \bibinfo {pages} {1231--1234}
  (\bibinfo {year} {1990})}\BibitemShut {NoStop}%
\end{thebibliography}%
\end{document}